# Neuroprotective Immunity by Essential Nutrient "Choline" for the Prevention of SARS CoV2 Infections: An in Silico Study by Molecular Dynamics Approach

Papia Chowdhury[1*], Pustak Pathak[2]

[1]Department of Physics and Materials Science and Engineering,
Jaypee Institute of Information Technology, Noida 201309, Uttar Pradesh, India

[2]Vishwa Bharati Public School, Arun Vihar, Noida, Sector 28, 201301, Uttar Pradesh, India

*Corresponding author: papia.chowdhury@jiit.ac.in

**Abstract:**

Prenatal COVID infection is one of the worst affected and least attended aspects of the COVID-19 disease. Like other coronaviruses, CoV2 infection is anticipated to affect fetal development by maternal inflammatory response on the fetus and placenta. Studies showed that higher prenatal choline level in mother's body can safeguard the developing brain of the fetus from the adverse effects of CoV2 infection. Choline is commonly used as food supplement. By virtual screening, molecular docking and molecular dynamics techniques, we have established a strong inhibitory possibility of choline for SARS 3CL[pro] protease which may provide a lead for prenatal COVID-19 treatment.

**Keywords:** COVID-19, SARS-CoV2, 3CL[pro], choline, RMSD, RMSF

# 1. Introduction

The whole world is now facing a pandemic situation raised from the COVID-19 disease caused by a member of coronavirus family of viruses, namely SARS-CoV2. Coronaviruses are well known from last few decades. These viruses have specific enveloped RNA which can be the reason for various respiratory illnesses like common cold to fatal pneumonia of varying severity [1]. In 1930s, it was first found in domestic poultry and it was recognized as cause of respiratory, liver, gastrointestinal, and neurologic diseases in animals [2]. There are seven categories of coronavirus, out of which four can only cause mild illness with symptom of common cold to the healthy human being. In very rare cases, these viruses can cause severe lower respiratory tract infections and pneumonia for infants, immune-compromised patients and older people. The rest three categories: SARS-CoV (identified in 2003), MERS-CoV (identified in 2012) [3] and SARS-CoV2 (identified in 2019) [4] can cause severe respiratory infections which can sometimes become fatal to the humans. We are already witnessing the devastating and deadly outbreaks of SARS-CoV2 which began in December 2019 from Wuhan, China [4, 5]. Within a few weeks after its appearance, the virus spread worldwide by showing symptom of respiratory illness (both acute and severe) caused by newly named disease COVID-19 caused by the so called novel coronavirus SARS-CoV2 [6]. In addition to usual respiratory problem, SARS-CoV2 can progress to acute respiratory distress syndrome (ARDS) and death, too. It's observed fact that the risk of death and\or serious symptoms due to COVID-19 increases in older people and in patients with other serious medical issues including obesity, diabetes, and heart, lung, kidney or liver disease [7]. The outcome of the above mentioned severity can progress to respiratory failure requiring mechanical ventilation, shock, multi organ failure, and death [8]. CoV2s are zoonotic pathogens in nature which usually appear in animals first. Due to the similarities of 79% and 96% for complete genome sequence recognition rates of SARS-CoV and bat SARS-CoV (SARSr-COV-RaTG13) it's suggested that bats may be the hosts of CoV viruses [9]. Similarly, for CoV2 there may exist original, intermediate and final hosts and so the disease may transfer from infected animal to human. For SARS-CoV2, person-to-person transmission is the easier way to spread the disease which can happen easily through infected secretions mainly via respiratory droplets, surface contaminated and possibly by aerosol transmission. Till now the research output concludes that the virus can easily be transmited by symptomatic, asymptomatic and presymptomatic patients. As per latest report, COVID-19 patients are mainly treated by providing supportive treatment only [10]. Around 200 FDA approved drugs and vaccines through clinical trials have been registered. Some investigational antiviral drugs like: Remdesivir, Favipiravir, Hydrochloroquine, Chloroquine, etc. have been in use for patients with severe symptoms [11, 12]. However, toxicities associated with these drugs have already created some major fatal health issues [11]. Alternatively, some Immunomodulatory therapy by including convalescent plasma also are in use but as per today, specific drugs and vaccines to fight against COVID-19 are yet to be discovered or are still under development. As a part of this development, some particular nutrient food supplements and herbal medicines which specifically have some antiviral, immunomodulatory activities against different viruses like influenza, HIV, other coronaviruses can be considered with the aim of promoting the use of dietary therapy, hard immunity and herbal medicine as an alternative of COVID-19 therapies [13,14]. A set of research works have already suggested that nutrient food supplements and herbs possess a potential antiviral ability, hard immunity against SARS-CoV2 [14].

| a. Receptor : CL$^{pro}$ Protease : 6LU7 | b. Inhibitor: Choline ($C_5H_{14}NO$) |
|---|---|
| 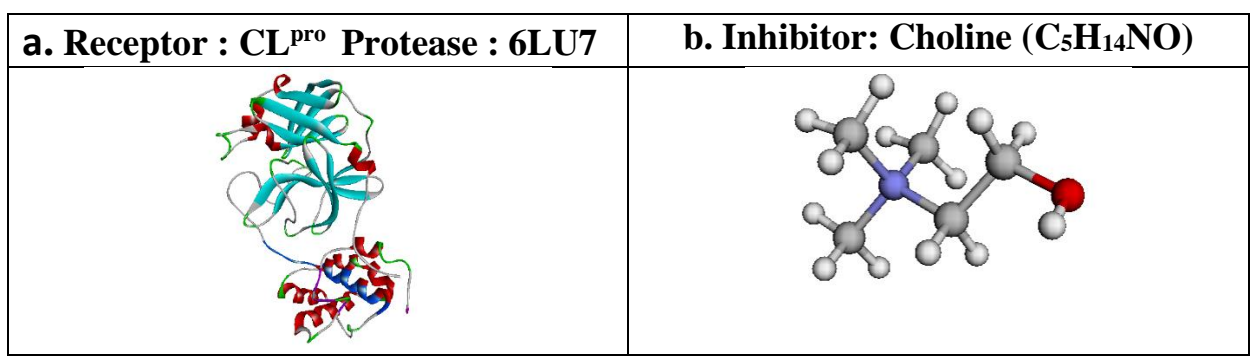 | |

**Figure 1. a)** Structure of receptor protein (6LU7). **b)**. Structure of Choline.

Previous pandemics due to different viruses have showed their effects on fetal developments which led to significant enhancement in number of infants with mental illnesses like schizophrenia, autism spectrum disorder and attention deficit disorder [15, 16]. As per report of "The Centers for Disease Control and Prevention (CDC)," prenatal COVID-19 infections can also develop its impact on fetal brain development by the effects of the maternal inflammatory response on the fetus and placenta [17]. We have to wait for some years for the supporting data regarding this impact from COVID-19. Current reported data from CDC based on Infant Behaviour Questionnaire Revised (IBQ-R) regulation [17, 18] suggest that no symptoms of viral infections are found after 3 months of birth on the children of COVID-19 infected mothers having high digestive choline level (≥7.5 μM) during the first 16 weeks of pregnancy. Some specific diets and food supplements are the source of choline for human body. Medical reports suggested that higher choline level in expecting mothers may protect fetal brain development from COVID-19, which definitely may support infant behavioral development though the mothers got COVID-19 infection in their gestation stage when the brain is being formed [17-20]. These findings have motivated us to work on a specific nutrient food supplement: *Choline* ($C_5H_{14}NO$) which is supposed to be used to protect the fetal brain development specifically during pregnancy [17]. Our *in silico* studies by virtual screening, molecular docking and molecular dynamics on choline as ligand and 3CL$^{pro}$ main spike protease of CoV2 as receptor have been able to establish that prenatal COVID-19 infections can be nullified by choline contained food supplements as they help to shield the fetal brain in mother's womb from the possible detrimental impact from COVID-19 (Figure 1a,b).

## 2. Materials and Methods:
## 2.1. Potential Inhibitor Choline

Choline is present in some dietary supplements/foods and belongs to the Vitamin B family. Choline is naturally found in many plant-based products like beans and cruciferous vegetables including dried nuts, whole grains, and seeds [21]. It's also found in animal-based products like poultry, dairy products, meat, eggs, and fish. It is one of the phytoconstituents extracted from herbal plants like *tinospora cordifolia*, *withania somnifera* [14]. Choline is endogenously produced in human liver as phosphatidylcholine. However, the amount of endogenously produced choline is not enough to meet our needs [22]. Choline plays a very significant role in forming of human genes, various metabolism process, and lipid transport. It helps the cell membranes to connect and communicate within the whole body. Choline is an important component of breast milk on which the fetal development of a child is dependent during the early birth stages. Medical studies reveal that proper intake of choline supplements by would be mothers can protect the infants from neurodegenerative diseases like Down syndrome, Alzheimer's diseases [23]. Choline is a direct precursor of Ach (acetylcholine) which is a neurotransmitter that controls a large number of autonomic, cognitive, and motor functions [23]. As per the medical studies, if choline are taken by the mother during pregnancy, this would make the newborn baby capable enough to mitigate the risks of CoV2 infections [17]. COVID-19 disease can destroy a child's respiratory organs and also hamper brain development. It's expected that choline can reduce the impact of the virus on the child's brain. The medical experts have relied on the fact that choline supplements will help the infants survive in a healthy manner though the mother get affected by the CoV2 infection. However, not much research justifying this hypothesis is conducted until now. Rigorous investigations are still needed to explore the connection between choline and SARS CoV2 infections and to determine whether choline can act as a potential inhibitor against the receptor protein CoV-2 main protease and to be used as potential drug so that choline supplements might benefit patients with COVID-19 disease. For the present study we have downloaded the structure of choline in PDB format from Drugbank (https://www.drugbank.ca/drugs) (Figure 1b). We have also done the proper virtual screening by evaluating their drug-likeness, pharmacokinetics and lipophilicity properties which are a set of guidelines for identification of potential drug compounds (Table 1). Some mostly industry applicable drug-likeness rules are: "Lipinski's rule", "MDDR-like rule", "Veber's rule", "Ghose filter", "Egan rule", "Muegge rule", etc [24-26]. SWISS-ADME software (https://www.swissadme.ch) has helped us in applying different virtual screening methods. For Molecular docking study, ligand choline has been saved in pdbqt format by Auto Dock Tools 1.5.6 [27].

| Properties of Choline($C_5H_{14}NO$) | | | | | | | | | |
|---|---|---|---|---|---|---|---|---|---|
| **Physicochemical Properties** | choline | *Lipophilicity* | choline | *Druglikeness* | choline | *Pharmacokinetics* | choline | **Medicinal Chemistry** | choline |
| *Molecular weight (gm/mol)* | 104.1708 | *Log $P_{o/w}$ (iLOGP)* | -2.41 | *Lipinski* | yes | *GI absorption* | Yes | PAINS | 0 alert |
| *Num. H-bond acceptors* | 1 | *Log $P_{o/w}$ (XLOGP3)* | -0.40 | *Veber* | yes | *BBB permeant* | No | Brenk | 1 alert |
| *Num. H-bond donors* | 1 | *Log $P_{o/w}$ (WLOGP)* | -0.32 | *Ghose* | Partly yes (2 violation) | *P-gp substrate* | No | Leadlikeness | Partially yes (1 violation) |
| *No of rotatable bonds* | 2 | *Log $P_{o/w}$ (MLOGP)* | -3.46 | *Egan* | yes | *Log $K_p$ (skin permeation)* | -7.22 cm/s | Mutagenicity | No |
| *Molar Refractivity* | 29.69 | *Log $P_{o/w}$ (SILICOS-IT)* | -0.57 | *Bioavailability score* | 137.87 | *Water Solubility* | -1.26 | Liver and cardio cytotoxicity | No |
| *Topological polar surface area TPSA ($Å^2$)* | 20.23 $Å^2$ | *Concensus Log Po/w* | -1.38 | *Synthetic accessibility (SA)* | 1.00 | *Log S (SILICOS-IT)* | 5.74e+00 | The Maximum Recommended Therapeutic Dose (MRTD) | 5.2mg/day |

**Table 1**. Physiochemical, Drug-likeness, pharmacokinetics and lipophilicity properties, Medicinal chemistry of choline.

## 2.2. Potential Target Protein Structure for SARS-CoV-2

Corona virus encodes large number of structural and nonstructural polyproteins. Among them some polyproteins become cleaved and transformed themselves into some mature nonstructural protein by protease 3CL$^{pro}$ which is a key to SARS-CoV enzyme [28]. CoV 3CL$^{pro}$ is responsible for controlling functions replication and transcription processes by its spike (S) glycoprotein [29]. S enters the host cell and from the viral surface it forms homo trimers. After entering it interacts strongly with the ACE2 (angiotensin-converting enzyme 2) receptor and replicates itself through some cyclic processes [30]. Targeting 3CL$^{pro}$ protease may constitute a valid approach for anti COVID drug designing. 3D structure of one of the 3CL$^{pro}$ like protease protein is reported by X-Ray crystallographic data (PDBID: 6LU7) [31] (Figure 1a). We have checked the inhibiting and binding possibilities of this natural 3CL$^{pro}$ like protease protein substrate in order to find the effectiveness of choline against COVID-19 in the present work. The structure of 6LU7 was downloaded from the "Protein Data Bank" (https://www.rcsb.org/) and has been prepared for further simulation purposes by Auto Dock and MG Tools of Auto DockVina software [32]. The inbuilt ligands were removed from 6LU7 using Discovery studio 2020 (*Dassault Systèmes BIOVIA*) [33] and the output structure was subsequently cleaned and saved in PDB format.

## 2.3. Methods: Molecular Docking, Molecular Dynamics, Binding Energy

Molecular docking which performs energy minimization and binding energy calculations, is one of the most applied simulation mechanisms used to predict the potential drug-target interactions. It can identify the best orientation/pose of ligand towards host protein. Docking algorithms with the help of some scoring functions are used to identify the suitable ligand conformation at energy minimized state [34]. Molecular docking can predict whether the ligand/drug is docked with the receptor protein/DNA or not and show the results in the form of rankings of docked compounds which are dependent on lower binding energy. The algorithm of AutoDock Vina with some best fitted configuration parameters (binding modes- 9, exhaustiveness – 8, energy difference- 3 kcal/mol, Gridbox center with coordinate x y, and z of residue position of the target protein) is used to do the docking-based studies on the suggested inhibitor onto the protease of CoV-2 [32]. The target protein is saved in pdbqt format by Auto Dock Tools [32] and ready to be used for docking. After simulation, the pose showing maximum nonbonded interactions, dipole moment with minimum binding affinity (kcal/mol), drieding energy, inhibition constant was chosen as the best ligand: protein complex structure. Another criteria of choosing best ligand: protein pose is identifying the types and number of bonding between them. The metabolite which makes maximum number of H-bonds, electrostatic bonds with the receptor protein mostly show better capability to form ligand: protein complex.

MD simulation results helped us to investigate the structural dynamics of protein: ligand interactions. LINUX based platform ''GROMACS 5.1 Package'' [35] with CHARMM36 all atom [36] and GROMOS43A2 force fields [37] are used for determination of various thermodynamic parameters like potential energy ($E_{pot}$), temperature (T), density progression (D), radius of gyration ($R_g$), root mean square deviation (RMSD) for backbone, root mean square fluctuation (RMSF) for protein $C_\alpha$ backbone, solvent accessible surface area (SASA), hydrogen bonds, interaction energies of the proposed ligand: protein complex. Choline was optimized by Gaussian 9.0 by Density Functional Theory (DFT) with the basis set 6.31G (d,p) [38, 39]. TIP3P water model and steepest descent algorithm has been used for energy minimization of the system in two steps each with varying time (1 ns – 10 ns) for 500,000 iteration steps with a cut-off up to 1000 kJmol$^{-1}$ for reducing the steric clashes. In first step, a boundary condition of constant number of particles (N), volume (V), and temperature (T) has been applied and in second step, it was constant NPT (particle numbers, pressure, temperature) under the pressure of 1 atmosphere and temperature 298K. For the nonbonded interactions, short range Lennard-Jones and Coulomb interaction have been used. After final step of each simulation, obtained trajectories and results were analyzed using the graphical tool Origin pro.

To compute the interaction free energies of protein: ligand complex ($\Delta G_{bind}$), we have used Molecular Mechanics Poisson-Boltzmann Surface Area (MMPBSA) method [40] sourced from the GROMACS and APBS packages. The model contains both repulsive and attractive components [41]. The snapshots at every 100 ps between 0 and 10 ns were collected to predict the binding energy. The binding free energy of the bound ligand: receptor complex in aqueous solvent can be expressed by following equations:

$$\Delta G_{bind,aqu} = \Delta H - T\Delta S \approx \Delta E_{MM} + \Delta G_{bind,solv} - T\Delta S \ldots\ldots\ldots\ldots\ldots..(1)$$

$$\Delta E_{MM} = \Delta E_{covalent} + \Delta E_{electrostatic} + \Delta E_{Van\ der\ Waals} \ldots\ldots\ldots\ldots\ldots.(2)$$

$$\Delta E_{covalent} = \Delta E_{bond} + \Delta E_{angle} + \Delta E_{torsion} \ldots\ldots\ldots\ldots\ldots\ldots.(3)$$

$$\Delta G_{bind,solv} = \Delta G_{polar} + \Delta G_{nonpolar}, \ldots\ldots\ldots\ldots\ldots\ldots\ldots\ldots\ldots.(4)$$

where $\Delta E_{MM}$, $\Delta G_{bind,solv}$, -$T\Delta S$ are the molecular mechanical energy changes in gas phase, solvation free energy change and conformational energy change due to binding, respectively. $\Delta E_{MM}$ is the combination of covalent, electrostatic and Van der Waals energy changes. Covalent energy is the combination of bond, angle and torsion. $\Delta G_{bind,solv}$ is separated into its polar and nonpolar contribution. The entropy term is approximated with a normal mode method using a few selected snapshots. For binding free energy calculation, the MMPBSA method usually begins after the MD simulation of the complex using the single-trajectory approach [42].

### 2.4. Computational Facility

For the MD simulation and GMM/PBSA computation works we have used Dell Gen9 system (8 Core I7 processors) with 16 GB of RAM with GPU NVIDIA MX130.

## 3. Results and Discussion

### 3.1. Analysis of Drug Likeness Properties of Choline

The inhibition properties of probe molecule choline have been studied through SWISS ADME software to determine physicochemical descriptors and to predict ADME (Adsorption, Distribution, Metabolism and Excretion) parameters: pharmacokinetic properties, drug like nature and medicinal chemistry friendliness [43]. From the data mentioned in Table 1 we can see that choline was satisfying the most prescribed virtual screening properties like Ro5 as it has molecular weight less than 500 g/mol, TPSA values less than 40 Å². H-bond donors ≤ 5, H-bond acceptor ≤ 10, synthetic accessibility count between less than 5 so that they can be synthesized easily. It also follows Veber rule that means it satisfies all the bioavailability conditions. Its pharmacokinetics validate its neuroprotective role for the brain development. All medicinal chemistry parameters suggests that choline has no liver cytotoxicity effect in terms or drug induced liver injury for humans, null mutagenicity reveals that choline can cause no abnormal genetic mutations leading to cancer. It has no cardio toxicity also. According to ADMET (https://vnnadmet.bhsai.org/) data the maximum recommended therapeutic dose is 5.2 mg/day.

### 3.2. Analysis of Molecular Docking Results

For receptor protein: 6LU7 and potential inhibitor: choline, docking result showed best 9 suitable pose structures for the choline: 6LU7 complex with their RMSD/ub (upper bound) and RMSD/lb (lower bound) variations. For pose 1 we obtained highest negative value of binding energy (-3.7 kcal/mol), lowest value of complex driending energy (104.198), and lowest value of inhibition constant ($k_i$ = 1.92

x10$^{-3}$ M) at room temperature (298 K) as the best fit structure of ligand: receptor complex (Figure 2, Table 2). We have repeated the docking simulation several times and got the variation in binding affinity for the best pose structure in between -3.7 kcal/mol to -3.4 kcal/mol. Usually, the dreiding energy values for individual protein and ligand are found to be different, but when they form complex the dreiding energy of the complex is found to be smaller than that of the individual protein and ligand. The complex structure having minimum dreiding energy corresponds to the most favourable structure [44]. In the same way for the best pose (1) structure the lowest $k_i$ value proved the higher affinity of choline towards receptor 6LU7. The computed $k_i$ value is far lower than the toxicity dose range of choline (Table 1) which validate the strong candidature of choline as a proposed drug for CoV2 infection. Maximum value of dipole moment (5.374 debye) for pose 1 also validate the possibility of better complexation between choline and 6LU7 which arises due to the existence of strong electrostatic interaction (attractive type) between 6LU7 residue GLU166 and positively charged nitrogen (N1) of choline. Again, the weak hydrogen bonded interactions have significant role in defining the stability of protein: ligand complex as larger the number of nonbonded interactions the more is the possibility of binding affinity in ligand: protein complexation and so the possibility of formation of complex structure [45]. Maximum number of weak interactions (both conventional hydrogen bonds and carbon hydrogen bonds) were observed for pose 1 of docked structure between 6LU7 (Residues: GLY143, LEU141, HIS163, PHE140, ASN142) and choline (atoms: O1, H15, C4, C2, C3) (Figure 2a,b, Table 2). So molecular docking results indicate that choline can be easily inhibited inside the favourable pocket of 6LU7 protein and can form a stable choline: 6LU7 complex by strong electrostatic bonding and weak hydrogen bonding between them. Existence of strong interaction between choline and 6LU7 have been established by MD simulation results in next section.

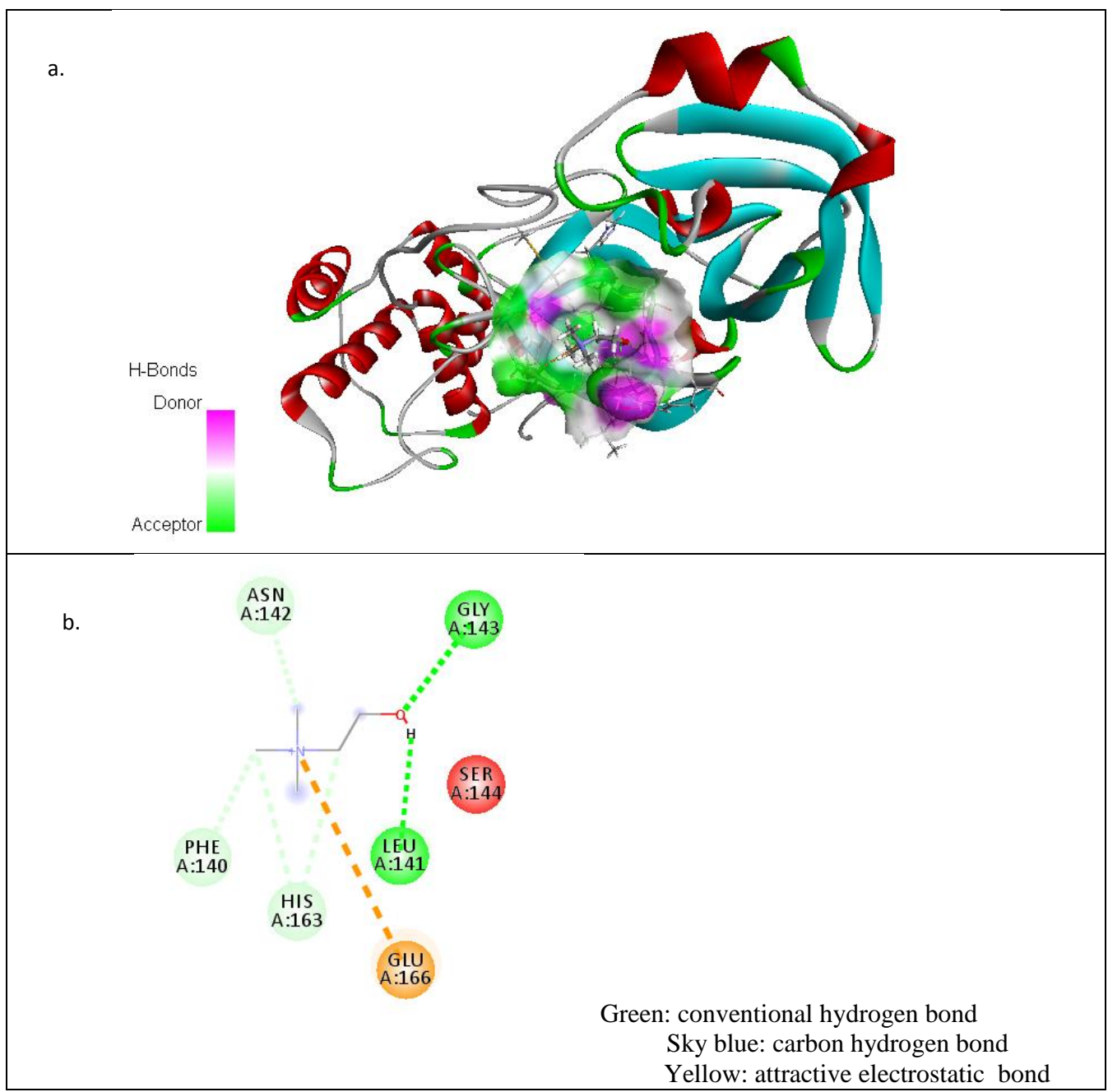

**Figure 2. a)** Donor: acceptor surface for hydrogen bonds for choline: 6LU7. **b).**Possible types of interactions in best pose structure obtained from molecular docking for choline: 6LU7.

| Ligand | Best Binding affinity (kcal/mole) | Hydrogen bonded interaction (donor: acceptor, distance in Å) [Type of bond] | Electrostatic interaction (ligand donor: proton acceptor, distance in Å) [Type of bond] | Dipole moment of ligand (debye) | Drieding energy between protein and ligand | Inhibition constant (M) |
|---|---|---|---|---|---|---|
| choline | -3.7 | (A:GLY143:HN - :PYA0:O1, 2.53816)[Conventional Hydrogen Bond]<br><br>(:PYA0:H14 - A:LEU141:O, 2.6499)[Conventional Hydrogen Bond]<br><br>(:PYA0:C4 - A:HIS163:NE2, 3.58031) [Carbon Hydrogen Bond]<br><br>(:PYA0:C2 - A:PHE140:O, 3.6244) [Carbon Hydrogen Bond]<br><br>(:PYA0:C2 - A:HIS163:NE2, 3.54655) [Carbon Hydrogen Bond]<br><br>(:PYA0:C3 - A:ASN142:OD1, 3.243) [Carbon Hydrogen Bond] | (::PYA0:N1- A:GLU166:OE2, 5.10169)[ Attractive Charge] | 5.374 | 104.198 | $1.92 \times 10^{-3}$ |

**Table 2**. Various parameters like binding affinity, hydrogen bonded interaction, electrostatic interaction, dipole moment, drieding energy, inhibition constant for best docked pose structure for choline: 6LU7 complex.

### 3.3. Analysis of Molecular Dynamics (MD) simulation Results

As per protocol of MD simulation, we have used the TIP3P water model to fulfill the solvation process in a cubic box of the current choline: 6LU7 complex by adding 4Na$^+$ external ions to maintain the neutrality of the present structure (Figure 3a). For MD simulation, used force field is one of the important parameter to estimate the intramolecular forces within the molecule and intermolecular forces between molecules which helps to calculate the potential energy of the system of atoms or molecules. The decomposition of the terms in the used force fields (CHARM36 and GROMOS43A2) is additive in nature in terms of energy as:

$$E_{total} = E_{bonded} + E_{nonbonded} \dots\dots\dots\dots\dots\dots\dots\dots\dots\dots\dots\dots(5)$$

$$E_{bonded} = E_{bond} + E_{angle} + E_{dihedral} \dots\dots\dots\dots\dots\dots\dots\dots(6)$$

$$E_{nonbonded} = E_{hydrogen\ bond} + E_{electrostatic} + E_{Van\ der\ Waals} \dots\dots\dots\dots\dots\dots(7)$$

$$E_{electrostatic} = E_{coulombic} + E_{lenard\ Jones} \dots\dots\dots\dots\dots\dots(8)$$

To quantify the strength of interaction between protein and ligand, nonbonded electrostatic interaction plays a vital part. For nonbonded interaction, we are interested mainly on short range electrostatic part. The whole MD simulation is performed under the varying time scale from 10000 ps to 100 ps. First, we have studied the stability of each structures (host and complex). We have optimized the probe systems by energy minimization process to get the minimum potential energy for both. For energetically stable structures, we obtained steady convergence of potential energy with negative energy minimum and maximum force value (Figure. 3b). For 6LU7 the computed average energy was -1.27x10$^6$ ±56.7 (kJ mol$^{-1}$) for E$_{pot}$, while for choline: 6LU7 it was -8.5 x 10$^5$ ± 38.77 (kJ mol$^{-1}$). Further to check the stability of the structures under NVT and NPT equilibrated ensembles, we have tried to compute T, D, P, V of the probe systems within a varying time trajectory 100ps – 10000 ps. Simulated progression data showed that temperature of each system (receptor, complex) reached to a stable value (298K) very quickly (within 100 ps) and maintained the stability throughout the whole simulation process for both applied force fields (cf. Figure 4a). Same type of stability has been observed for density, pressure, volume throughout the varying time scale (Inset of Figure 4a). After running the whole simulation process we have computed several thermodynamics parameters of choline: 6LU7 complex in comparison to bare 6LU7 to understand the possible conformational arrangements during complexation throughout the whole time trajectory. The parameters are RMSD, RMSF, nonbonded interaction energy, hydrogen bonds, R$_g$, etc. (Table 3) [46]. RMSD is used to check the stability of host protein due to ligand binding with respect to the reference backbone structures of protein. Figure 4b has shown the RMSD values of carbon backbone for choline: 6LU7 complex with respect to 6LU7 for the time scale trajectory up to 1 ns in 3D view (Figure 4b). For better understanding a 2D diagram is also provided in the inset of Figure 4b for larger time trajectory 1 ns-10 ns. It is observed that RMSD for complex showed a variation between 0.14-0.26 (± 0.01) nm compared to 6LU7 variation 0.12- 0.18 (± 0.01) nm which indicate the possibility of structure fluctuation during choline binding. The larger fluctuation in complex structure mainly arises between 7.5 ns – 10 ns time scale. Still the closeness between the average RMSD values of complex (0.18 nm) with host 6LU7 (0.16 nm) validated the stable docked choline: 6LU7 formation. Similarly for RMSF, residue fluctuations of 6LU7 in complex were found to be very small with respect to the 6LU7 in terms of C$_\alpha$ carbon. Closeness and less fluctuations in RMSF values suggest that choline: 6LU7 docked complex structure has not effected the host protein structure which remained mostly unaltered during simulation (supporting document SD1).

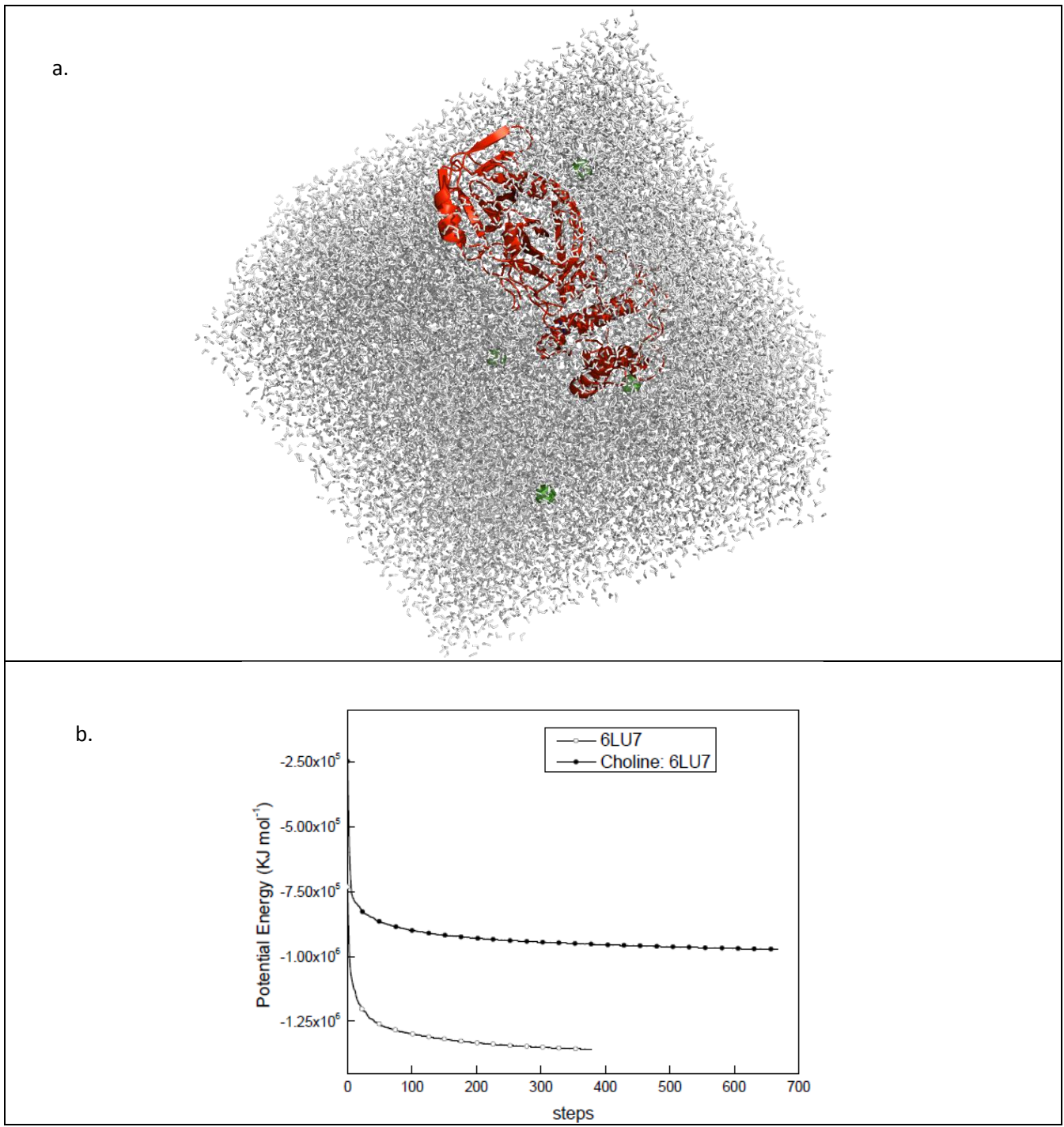

**Figure 3. a)** Solvated and neutralized choline: 6LU7 complex system in presence of water environment Na$^+$ ions. **b).** Potential energy surface for optimized geometries of choline: 6LU7 complex and bare 6LU7.

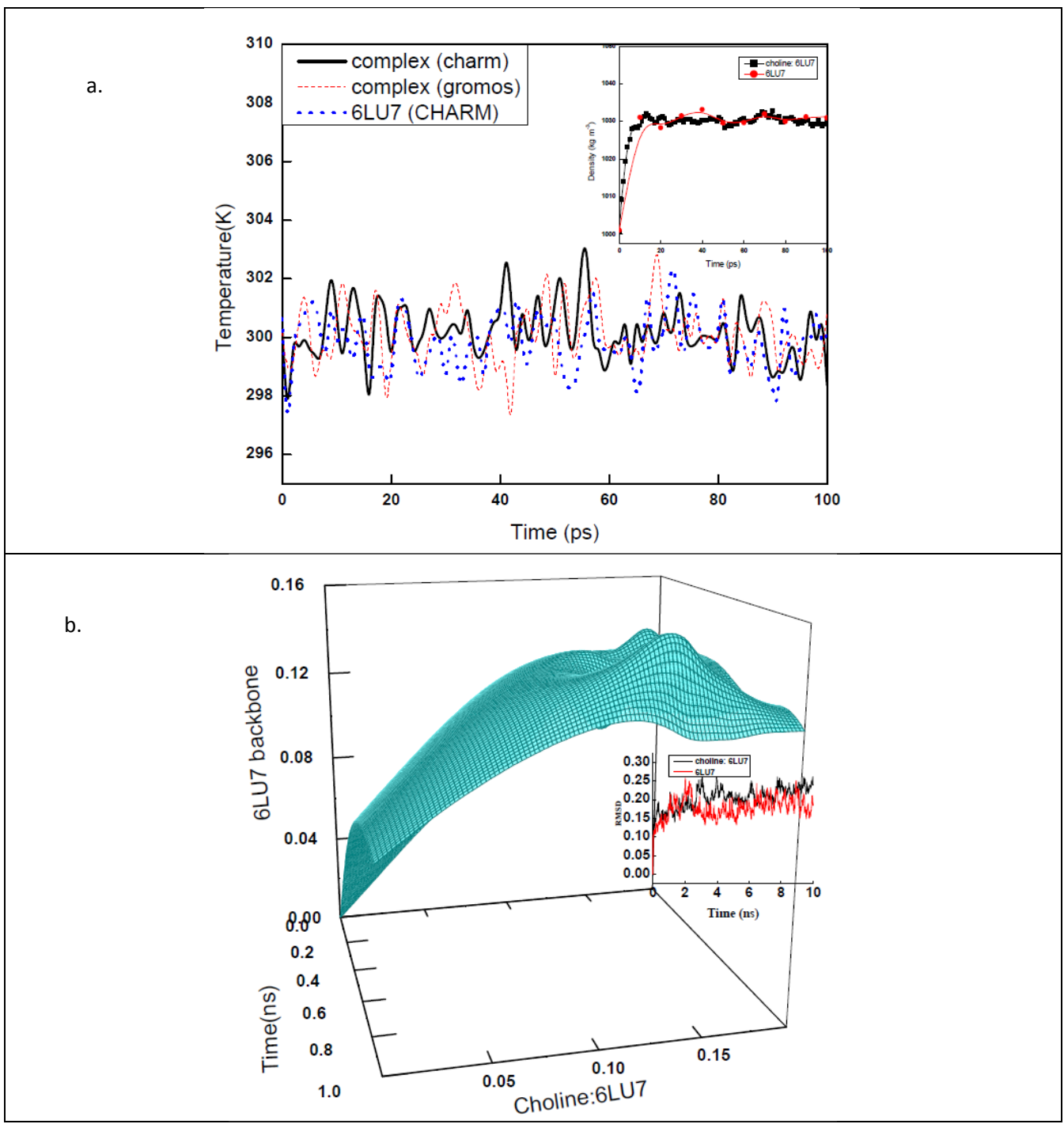

**Figure 4.a)** Temperature and density (inset) progression data for bare 6LU7 and choline: 6LU7 complex in water environment in CHARM 36 and GROMOS force fields. **b)** Root mean square Deviation (RMSD) backbone graphs of 6LU7 in its bare state and in complex with choline (3D view upto 1 ns, 2D view upto 10 ns).

| Serial no | Parameter | Bare 3CL$^{pro}$ protease (6LU7) | | 6LU7 + choline complex | |
|---|---|---|---|---|---|
| | | Mean | Range | Mean | Range |
| 1. | SR Columbic Interaction energy (kJ mol$^{-1}$) | NA | NA | -50.776± 20.12 | -70 - -10 |
| 2. | SR LJ Interaction energy (kJ mol$^{-1}$) | NA | NA | -95.26± 22.10 | -30 - -120 |
| 3. | Average Interaction Energy energy (kJ mol$^{-1}$) | NA | NA | -160.95± 20 | NA |
| 4. | RMSD (nm) | 0.16± 0.01 | 0.12 – 0.18 | 0.18± 0.01 | 0.12 – 0.26 |
| 5. | Inter H-Bonds | NA | NA | 2.5 | 0 - 6 |
| 6. | Radius of gyration | 2.25± 0.01 | 2.25 – 2.26 | 2.25± 0.01 | 2.25 – 2.26 |
| 7. | SASA (nm$^2$) | 22 | 19 - 26 | 19 | 16-20 |
| 9. | Potential Energy (kJ mol$^{-1}$) | -1.27x10$^6$ ±56.7 | -7.3 x10$^5$ - -1.3 x10$^6$ | -8.5 x 10$^5$ ± 38.77 | -5.4x10$^5$ - -1.0x10$^6$ |
| 10. | Binding Energy (ΔG) (kJ mol$^{-1}$) | NA | NA | -33.497 +/- 27.406 | NA |
| 11. | Van der Waal Energy (ΔE$_{vdW}$) (kJ mol$^{-1}$) | NA | NA | -34.656 +/- 24.958 | NA |
| 12. | Electrostatic Eenergy (ΔE$_{elec}$), (kJ mol$^{-1}$) | NA | NA | -7.380 +/- 11.506 | NA |
| 13. | Polar solvation Energy (ΔE$_{polar}$) (kJ mol$^{-1}$) | NA | NA | 12.790 +/- 28.852 | NA |
| 14. | SASA Energy (ΔE$_{apolar}$ kJ mol$^{-1}$) | NA | NA | -4.251 +/- 3.148 | NA |

**Table 3.** Statistical data obtained by Molecular dynamic simulations for 6LU7 in its bare state without any ligand and in the complex state with ligand choline.

The radius of gyration ($R_g$) of a protein ligand complex or basic backbone protein gives us the hint of compressed nature of the probe system [47]. Variation of $R_g$ values over the whole time trajectory showed that the host 6LU7 and choline: 6LU7 complex showed a quite stable and compressed structures without lacking any major expansion/contraction throughout the simulation period. Both complex and base protein structures have reported $R_g$ values between 2.25-2.26 nm with an average value of 2.25±0.01 nm (SD2). Currently mostly used anti-COVID-19 drugs are Remdesivir, Favipiravir, hydrochloroquine. In Remdesivir the average $R_g$ value is reported to be 2.2 ±0.1 nm [48]. Favipiravir and hydroxychloroquine are also expected to show similar compactness due to their similar type of binding affinity towards 3CL$^{pro}$ protease as per Ref. [49]. The value of Remdisivir exactly matches with the average $R_g$ value obtained for choline. This justifies the candidature of choline as a proposed CoV2 inhibitor. Similarly, SASA measures the area of receptor exposure to the solvents during the simulation process. In present study for 6LU7, SASA value was obtained between 19-26 nm$^2$ with a mean value of 22 nm$^2$ whereas for choline: 6LU7 complex, it was observed between 16-20 nm$^2$ with a mean value of 19 nm$^2$ (SD3). The closeness of the observed SASA values of both bare 6LU7 and it complex justified that ligand binding does not affect the folding of the receptor protein very much.

Though weaker than ionic and covalent interaction, intermolecular hydrogen bonded interaction is a predominant contributor for complex formation. 3.5Å cut-off condition has been used to find number of hydrogen bonds. In the present study we have observed a variation of number of hydrogen bonding between 0 to 6 throughout the whole time trajectory with an average value of ±2.6 (SD4). Obtained numbers perfectly matches with the molecular docking results and further validate the stability of choline: 6LU7 complexation. Short-range nonbonded interaction energy used to quantify the strength of the interaction between components of a complex. For the present case we have presented the variation of short range Coulombic interaction energy and short range Lennard-Jones interaction energy versus time (0- 10000 ps) for choline: 6LU7 complex and presented in 3D Figure 5a and Table 3. Energy values showed a stable nature up to 7000 ps of time scale data. Computed data revealed the less effect of binding affinity due to Coulombic interaction energy (-50.776± 20.12 kJ mol$^{-1}$ ) compared to Lennard-Jones interaction energy (-95.26± 22.10 kJ mol$^{-1}$ ). Time resolved variation of energy values are also found to match perfectly with the RMSDs of the complex for full time scale region.

To establish the stronger ligand binding affinities towards receptor we have applied the MM/PBSA method which deals with net free binding energy (ΔG) change as a sum of comprehensive set of energies of individual components. Molecular docking only suggest the binding energy of the complex while ΔG indicates the nonbonding interaction energies of the binding region for the complex formation. The average MM/PBSA free binding energy (ΔG$_{bind}$) of choline in complex was obtained as -33.497 kJ/mol which is good enough due to its well binding affinity towards SARS-CoV2 protease. The Van der Waal energy (E$_{vdw}$) component (-34.656 kJ/mol) also showed very good binding affinity of choline towards 6LU7 whereas electrostatic energy (E$_{electrostatic}$) did not show any significant role (-7.380 kJ/mol) in the complexation process. The variation of all energies needed for ΔG (energy in vacuum, polar and nonpolar solvation energies) is described in Figure 5b. Figure 5c shows the average free ΔG$_{bind}$ in the whole time trajectory. All the MD simulation results reported in this work and summarized above validate that choline can make an impressively stable complex with SARS-Cov2 protein after binding to the active sites of this 3CL$^{pro}$ protease.

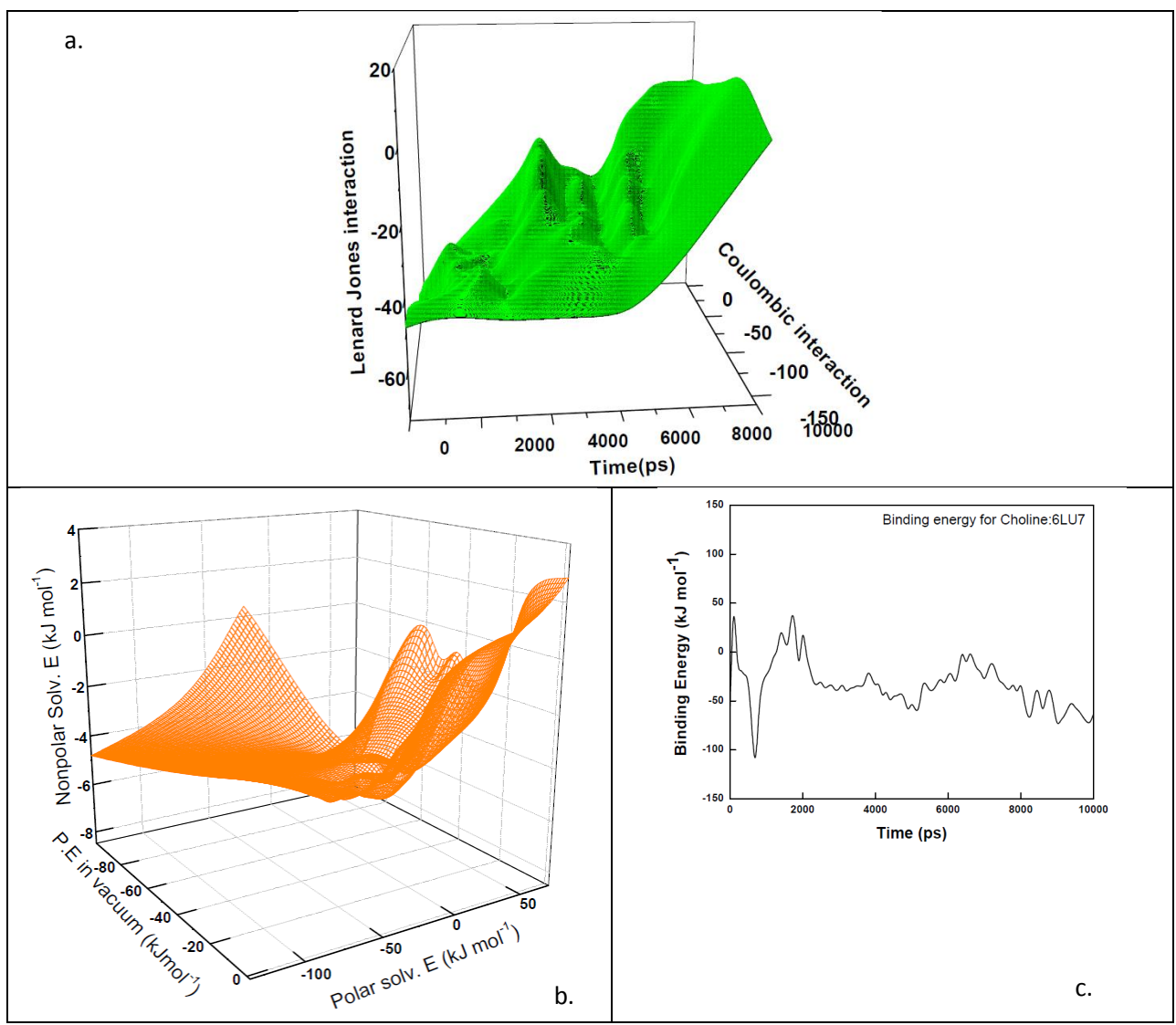

**Figure 5.** For choline: 6LU7 complex: **a).** Variation of Coulombic interaction energy and Lenard Jones interaction energy, **b).** Relation between Energy in vacuum, Polar solvation energy, Nonpolar solvation energy, **c)**. Variation of total Free binding energy with respect to full time trajectory (0 ps – 10000 ps).

## 4. Conclusion

CoV2 infection is anticipated to affect fetal development in a manner similar to other respiratory viruses which are known to be responsible for the maternal inflammatory response on the fetus and placenta. Earlier studies have established that if would be mother gets infected by a respiratory coronaviruses during early pregnancy, higher prenatal choline level in her body can safeguard the developing brain of the fetus from the adverse effects of CoV2 infection. There are many natural recourses of choline (e.g., vegetables, egg, milk, whole grains, herbal plants). Our ADME analysis (physiochemical, pharmacokinetics, drug likeness, medicinal chemistry) have found a strong inhibitory possibility of choline for SARS-CoV2 protease 3CL$^{pro}$. Molecular docking results with strong binding binding affinity, lowest inhibition constant and existence of hydrogen bonded interaction have established the possibility of existence of choline: 6LU7 complex structure. Various thermodynamic parameters ($E_{pot}$, T, V, D, $R_g$, SASA energy, interaction energies, $\Delta G_{bind}$) obtained by Molecular dynamics simulation have validated the complexation between choline and CoV2 protein (6LU7). Summarizing all simulated output and interpreting their analysis have helped us to establish choline as a strong candidate to be used as a potential inhibitor for SARS-CoV2. We believe that our present in silico study would provide a lead for the prenatal drug development for the treatment of COVID-19. Still to efficiently target CoV2 infections, production of high quality concentrated choline extract and in vivo, in vitro experimental validation of the present work and clinical studies are defensible.